\documentclass[letterpaper,journal]{IEEEtran}
\usepackage{amsmath,amsfonts}
\usepackage{algpseudocode}
\usepackage{algorithm}
\usepackage{array}
\usepackage[caption=false,font=normalsize,labelfont=sf,textfont=sf]{subfig}
\usepackage{textcomp}
\usepackage{stfloats}
\usepackage{tikz}

\usetikzlibrary{decorations.pathreplacing}
\usetikzlibrary{arrows.meta}
\usepackage{xcolor}
\usepackage{url}
\usepackage{verbatim}
\usepackage{booktabs}
\usepackage{tabularx}
\usepackage{caption}
\usepackage{graphicx}
\usepackage{makecell}
\usepackage{hyperref}
\usepackage{cite}
\begin{document}

\title{ Delay Time Characterization on FPGA: A Low Nonlinearity, Picosecond Resolution Time-to-Digital Converter on 16-nm FPGA using Bin Sequence Calibration }

\author{Sunwoo Park, Byungkwon Park, Eunsung Kim, Jiwon Yune, Seungho Han and Seunggoo Nam

\thanks{Sunwoo Park, Byungkwon Park, Eunsung Kim, and Jiwon Yune are with the Division of Quantum Technology, SDT Inc., Seoul 06211, South Korea. \\ \\ \indent Seungho Han and Seunggoo Nam are with Korea Electronics Technology Institute (KETI), Seongnam-si, South Korea \\(e-mail: sunpark@sdt.inc; bkpark@sdt.inc; eskim@sdt.inc; jwyune@sdt.inc; shhan@keti.re.kr; senggu@keti.re.kr). 
}
}

\markboth{}%
{Shell \MakeLowercase{\textit{et al.}}: A Sample Article Using IEEEtran.cls for IEEE Journals}


\maketitle

\begin{abstract}
We present a Time-to-Digital Converter (TDC) implemented on a 16 nm Xilinx UltraScale+ FPGA that achieves a resolution of 1.15 ps, RMS precision of 3.38 ps, a differential nonlinearity (DNL) of [–0.43, 0.24] LSB, and an integral nonlinearity (INL) of [–2.67, 0.15] LSB. This work introduces two novel hardware-independent post-processing techniques — Partial Order Reconstruction (POR) and Iterative Time-bin Interleaving (ITI) — that significantly enhance the performance of FPGA-based TDCs.  POR addresses the missing code problem by inferring the partial order of each time bin through code density test data and directed acyclic graph (DAG) analysis, enabling near-complete recovery of usable bins. ITI further improves fine time resolution by merging multiple calibrated tapped delay lines (TDLs) into a single unified delay chain, achieving scalable resolution without resorting to averaging.  Compared to state-of-the-art FPGA-based TDC architectures, the proposed methods deliver competitive or superior performance with reduced hardware overhead. These techniques are broadly applicable to high-resolution time measurement and precise delay calibration in programmable logic platforms.

\end{abstract}


\section{Introduction}
\IEEEPARstart{H}{igh} resolution time interval measurements are fundamental in a wide range of fields, including LiDAR systems~\cite{lidar2}, medical imaging modalities~\cite{biomedical_imaging}, space science~\cite{space_science}, nuclear and particle physics~\cite{pnp1,pnp2}, positron imaging tomography~\cite{PET} quantum information processing~\cite{QI} including Quantum Key Distribution (QKD)~\cite{ref_QKD}, and quantum optics~\cite{ref_quantum_optics}. One method for high-resolution timing is the Time-to-Amplitude Converter (TAC)~\cite{ref_TAC}, which converts a time interval into a voltage amplitude for subsequent digitization.

Although TACs offer high precision and low integral nonlinearity, their measurement range is often limited, and their analog nature makes integration with digital systems less straightforward. This has led to the increasing adoption of time-to-digital converters (TDCs), which can offer compact, scalable, and fully digital timing solutions.  A TDC can be constructed from a Vernier Delay Line (VDL) and a Tapped Delay Line (TDL).  A TDL TDC has been the mainstream since it can easily be constructed from field-programmable gate array (FPGA) devices.  However, the conventional TDL TDCs suffer from limitations in resolution, nonlinearity, and bin uniformity, particularly when constrained by clock speeds and process variation. TDCs can also be implemented in an application-specific integrated circuit (ASCI) as well as on a FPGA. An ASCI TDC typically has better resolution and nonlinearity than a FPGA TDC and has analog calibration feasibility~\cite{Wu:2008zzr}.  However FPGA TDCs are more widely used in scientific research, thanks to their large flexibility, low development cost, and short development cycle.  With easy access and usefulness for product developments, FPGA TDCs have been the mainstream electronics.


The operating principle of a TDC is to subdivide the time axis into finer intervals beyond the resolution afforded by a high-frequency sampling clock. In modern FPGA-based systems, feasible clock frequencies are typically limited to a few hundred megahertz, corresponding to time resolutions on the order of nanoseconds. To achieve finer granularity, TDCs employ spatial delay lines that propagate an input signal through a chain of logic elements. The position of the signal along the delay line at the moment of clock arrival is sampled by discrete elements known as time bins, enabling fine time measurements.

The effective length of the delay line is constrained by the clock period, and the achievable resolution improves with faster signal propagation—since identical spatial separations then correspond to smaller temporal intervals.

A fundamental challenge in TDC implementation lies in the variation of delay paths among the time bins. At picosecond - level resolution, variations in signal propagation and clock skew between bins become significant. This results in time bins sampling the input at nonuniform offsets, leading to unordered bin sequences and ambiguous 0–1 transitions in the sampled thermometer code. Consequently, certain time bins may be skipped entirely, degrading both the accuracy and effective resolution of the TDC.

Beyond resolution, the three key figures of merit for a TDC are: resolution, accuracy, and nonlinearity.  The first two - resolution and accuracy - together characterize the system's ability to reliably distinguish short time intervals.  Several state-of-the-art techniques have been developed to improve these metrics, including the Wave Union(WU) method~\cite{Wu:2008zzr,fine_high_DL}, multi-chain averaging~\cite{Multi-chain1}, and the sub-TDL method~\cite{Sub-TDL1}.  The WU method injects patterned pulse trains to generate additional transitions, effectively subdividing wide time bins into finer and more uniform segments.  Multi-chain averaging enhances the measurement precision and makes use of time bins across different TDLs by averaging fine time outputs from the multiple distinct TDLs.  The sub TDL method strategically assign the multi-TDLs in multi-chain averaging, in a way that no time bins are missing and wasted.  

The third metric, nonlinearity, refers to how uniformly the time bins are distributed in the hardware implementation.  Due to clock jitter and circuit-level variation, repeated single-shot measurements may map to multiple neighboring bins rather than a single bin, thereby degrading measurement fidelity. A low nonlinearity ensures a consistent fine time resolution along the entire delay line and supports reliable statistical interpretation of results.  To mitigate nonlinearity, calibration methods such as bin-by-bin calibration~\cite{sensors} and bin-width calibration~\cite{sensors,embedded_bin_width_cal} are employed.  Additionally, techniques like multi-phasing~\cite{dual_phase,embedded_bin_width_cal} have been proposed to reduce integral nonlinearity (INL) across the delay line, thereby improving overall performance (see Section~\ref{subsec:Bin_width_calibration}).

In order to enhance the resolution and accuracy, this work introduces two novel post-processing techniques that can be applied on any hardware.  These techniques can not only advance the implementation of TDCs on FPGAs, but also offer general utility for precision digital design:

\begin{enumerate}
    \item \textbf{Partial Order Reconstruction (POR):} A calibration framework capable of experimentally characterizing sub-picosecond delay variations (demonstrated up to 0.2~ps) across FPGA routing elements. POR identifies and corrects missing codes by reconstructing the correct temporal ordering of time bins. It enables fine-grained delay calibration with minimal hardware overhead and is theoretically applicable to any FPGA architecture.
    
    \item \textbf{Iterative Time-bin Interleaving (ITI):} A post-calibration technique that merges multiple delay lines based on the timing information recovered via POR. ITI allows for scalable resolution enhancement and supports the construction of high-precision TDCs with arbitrarily fine bin widths.
\end{enumerate}

\section{Overall Architecture}

\begin{figure*}[!htbp]
  \centering
  \includegraphics[width=0.98\linewidth]{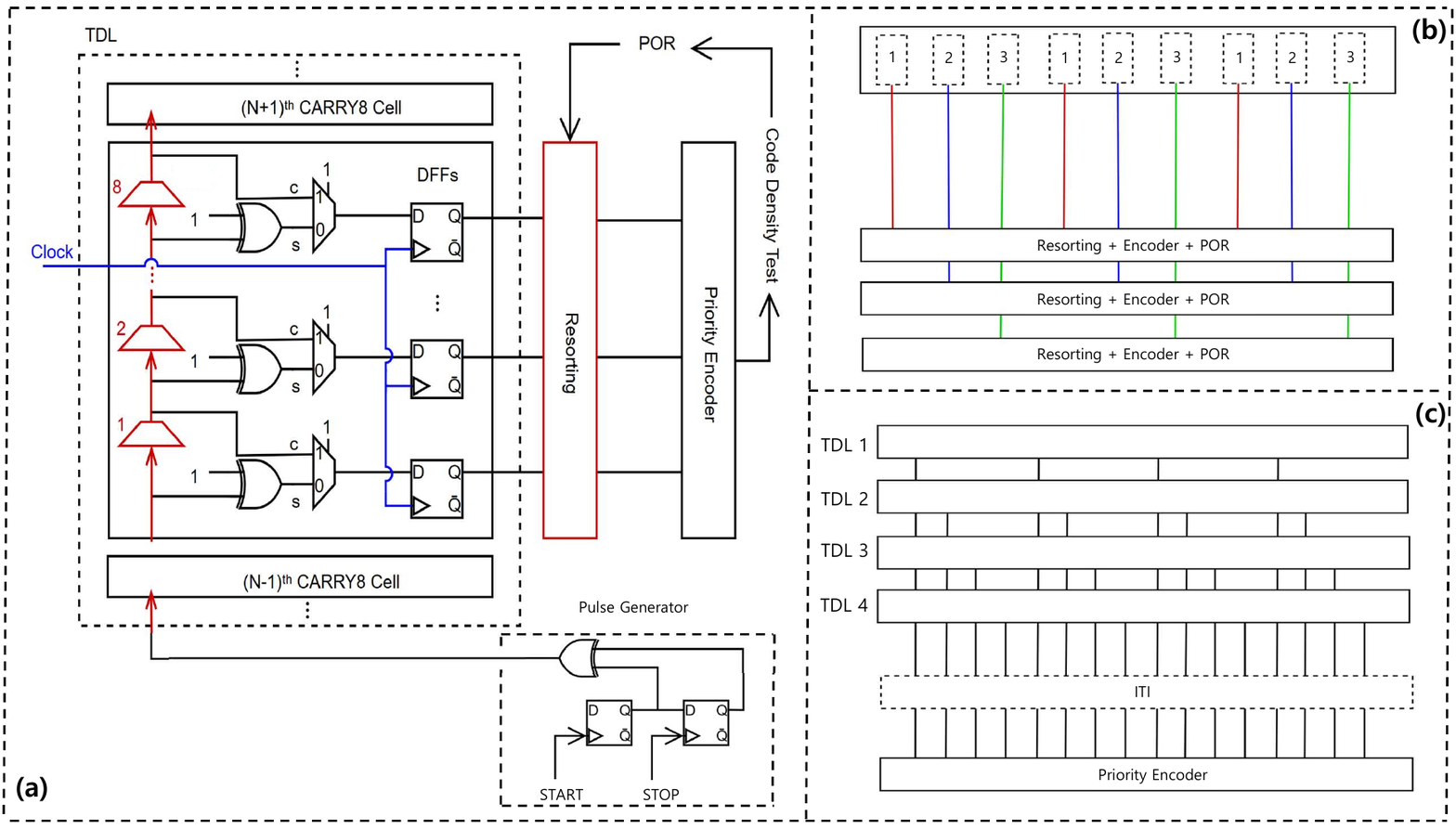}
  \vspace{-8pt}
  \caption{TDC architecture overview. (a) Functional block diagram. The input pulse is shaped with a fixed duration defined by the Start and Stop signals. The time bins (D flip-flops) are sorted via POR (see Section~\ref{subsec:POR}) based on code density test results. (b) Illustration of the $Z_3$ grouping used in POR. (c) Iterative Time-bin Interleaving (ITI) framework to combine multiple delay lines according to their calibrated bin times (see Section~\ref{subsec:ITI}).}
  \label{fig:TDC_design}
\end{figure*}

The TDC developed in this work is implemented on a 16\,nm FinFET-based Xilinx UltraScale+ FPGA. An overview of the architecture is shown in Figure~\ref{fig:TDC_design}. The implementation in this work uses only the `C' taps within each CARRY8 primitive as time bins, and operates with a 250\,MHz sampling clock.

To capture the signal position along the delay line , a series of D flip-flops (DFFs) are arranged along the delay line to form discrete time bins. When the sampling clock hits, each DFF samples the signal’s presence, producing a thermometer code (e.g., \texttt{11110000} in an ideal case without missing codes; see Section~\ref{subsec:POR}). This thermometer code (TC) is then converted into a one-hot code (OHC) by detecting the first 1--0 transition. The OHC is subsequently encoded into a binary value, representing the fine time output of the TDC.

To convert the binary output into an accurate time value, a calibration process known as the code density test is required. This involves feeding the TDC with asynchronous or uniformly distributed random input pulses. Over many samples, the relative frequency of each OHC reveals the effective width of its corresponding time bin. In this research, over at least 5 million shots of random pulses have been fed for each code density test, which only takes a few seconds to perform. By accumulating the bin widths sequentially, a fine time scale can be reconstructed. This bin-to-time map is stored in the processing system (PS) of the FPGA. In turn, each TDC output code can be accurately translated into its corresponding fine time value. Combined with the coarse counter tracked by the ARM processor embedded within the FPGA, this enables time measurements with both high resolution and extended dynamic range.  Then the full-resolution timestamp is transmitted externally via a Gigabit Ethernet interface. 

To mitigate missing codes from the bin ordering, the delay line is calibrated using the POR technique (Section~\ref{subsec:POR}). POR reorders the time bins at the encoder level based on experimental delay characterization, without modifying the underlying routing or logic.

After calibration, multiple TDLs are merged to form a single interleaved delay line through the ITI method (Section~\ref{subsec:ITI}). This merging process leverages the time bin information recovered by POR to enhance resolution beyond the limits of a single TDL.

\subsection{Bin Sequence Calibration: \\ Partial Order Reconstruction (POR) \label{subsec:POR}}

\begin{figure}[!htbp]
    \centering
    \resizebox{1.0\linewidth}{!}{\vspace{0.5cm}

\begin{tikzpicture}[x=1cm, y=1cm, font=\footnotesize]
\tikzstyle{bin} = [draw=black, minimum width=1cm, minimum height=0.8cm, align=center]

\node[anchor=west] at (0.5,1.0) {\textbf{(a) Perceived vs Actual}};

\node[anchor=west] at (8.9,0.0) {\textbf{Perceived}};
\foreach \i [count=\x from 1] in {1,2,3,4,5,6,7,8} {
    \node[bin] (perceived\i) at (\x,0) {\i};
}

\node[anchor=west] at (8.9,-1.2) {\textbf{Actual}};
\foreach \i [count=\x from 1] in {1,2,3,5,4,6,7,8} {
    \node[bin] (actual\i) at (\x,-1.2) {\i};
}

\draw[->, thick, >=Stealth] (perceived4.south) -- (actual4.north);
\draw[->, thick, >=Stealth] (perceived5.south) -- node[midway, right=5pt] {\scriptsize swapped} (actual5.north);

\node[anchor=west] at (0.5,-2.2) {\textbf{(b) Pulse Trains}};

\draw[thick, blue!50, >=Stealth]
  (0.5,-2.6) -- (4,-2.6) 
  -- (4,-3.0) 
  -- (8.5,-3.0);
\node[anchor=west, blue!60!black] at (8.7,-2.6) {\scriptsize (i) Pulse train 1};

\draw[thick, red!50, >=Stealth]
  (0.5,-3.4) -- (5,-3.4)
  -- (5,-3.8)
  -- (8.5,-3.8);
\node[anchor=west, red!60!black] at (8.7,-3.4) {\scriptsize (ii) Pulse train 2};

\node[anchor=west] at (0.5,-4.8) {\textbf{(c) Reconstructed Sequences}};

\node[left, red] at (0.3,-5.8) {\scriptsize (i)};
\node[left, blue] at (0.3,-7.1) {\scriptsize (ii)};

\foreach \i [count=\x from 1] in {1,2,3,4,5,6,7,8} {
    \pgfmathsetmacro{\fillcolor}{
        ifthenelse(\i==1 || \i==2 || \i==3 || \i==5, "blue!15", "white")
    }
    \node[bin, fill=\fillcolor] at (\x,-5.8) {\i};
}
\node[anchor=west] at (8.7,-5.8) {\textcolor{black}{$\Rightarrow$ OHC = 3}};

\foreach \i [count=\x from 1] in {1,2,3,4,5,6,7,8} {
    \pgfmathsetmacro{\fillcolor}{
        ifthenelse(\i==1 || \i==2 || \i==3 || \i==4 || \i==5, "blue!15", "white")
    }
    \node[bin, fill=\fillcolor] at (\x,-7.1) {\i};
}
\node[anchor=west] at (8.7,-7.1) {\textcolor{black}{$\Rightarrow$ OHC = 5}};

\end{tikzpicture}}
    \caption{Illustration of missing code formation due to bin sequence mismatch. 
    (a) A mismatch between the perceived and actual ordering of time bins results from delay variation. 
    (b) Two signals transitioning at the fourth (Blue) and fifth (Red) time bins. 
    (c) Resulting OHCs from the two signals. Time bin 4 is never selected as the OHC output and is thus effectively skipped, resulting in a missing code.}
    \label{fig:missing_code}
\end{figure}
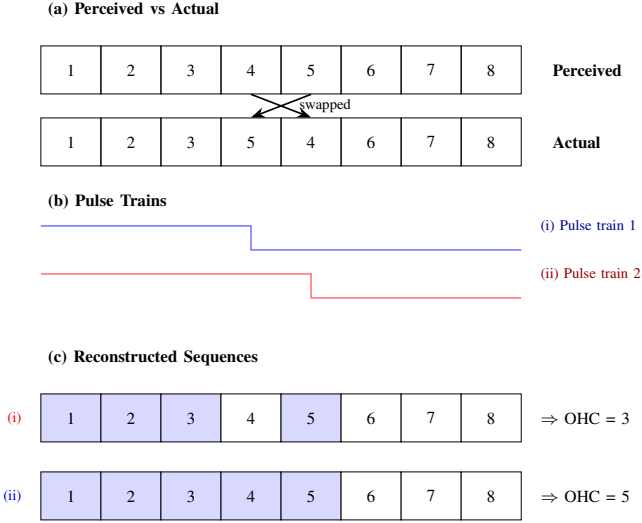

When operating at resolutions on the order of a few picoseconds, discrepancies between the perceived and actual bin sequences in FPGA-based TDCs become inevitable. Variations in bin-specific path delays, combined with clock skew, make it practically impossible to predict the bin structure with perfect accuracy in advance. This misalignment between expected and actual bin positions gives rise to missing codes — certain time bins are effectively skipped during fine time measurement (see Fig.~\ref{fig:missing_code}).  At the priority encoder, the 1-0 transition point is taken to be the bit before which a certain number of 1s repeat in the TC. 

Throughout this paper, we refer to bins that are not missing as 'tapped', and those that are missing as 'untapped'. The presence of missing codes degrades both the effective resolution and linearity of a TDC ($DNL = -1$).


This paper proposes a direct post-processing approach to calibrating the bin sequence by explicitly inferring and correcting the missing codes.  This method achieves fine time resolution while eliminating missing codes, thereby reducing hardware overhead. As such, it offers DFF resource efficiency in TDL implementation similar to that of the sub-TDL technique~\cite{Sub-TDL1}, although only a single priority encoder is needed using this method.  This method can be used in parallel with existing state-of-the-art techniques, such as the on-line bin-width calibration in \cite{embedded_bin_width_cal} which requires no missing codes.  The key insight is that the presence and positions of missing codes after a code density test conversely indicates 'partial orders' of the time bins. 

Consider a bin such that all preceding bins and itself are tapped (alternatively, the first bin after which bin sequence calibration will be performed).  We designate this as bin 1, without loss of generality.  We define the 'bridge' as the most recent bin whose position has been established in the ordering.  Initially, bin 1 is the bridge.  For bin 2 to be tapped, the following condition must hold: the next bin (bin 3) must not precede the bridge in the actual ordering.  By the same rule, since bin 1 is assumed to be tapped, bin 2 must appear after bin 1.  By definition, this now makes bin 2 the new bridge.  Now suppose bin 3 were to precede bin 2, such that the actual ordering is $(1,3,2)$.  In this case, a code density test would result in signals transitioning at both bin 1 and bin 3 yielding bin 1, while transitions at bin 2 would yield bin 3 - causing bin 2 to be a missing code.  Alternatively, if the actual sequence could be $(3,1,2)$ to produce the same missing code.  If, however, bin 2 is tapped, this confirms that the sequence is $(1,2,3)$.  If bin 2 is untapped, bin 3 could be located anywhere prior to bin 2.  

This observation motivates the following principle:
\begin{center}
    \textit{For a bin to be tapped, its next bin should NOT precede the current bridge, as determined from partial order deductions up to that point.}
\end{center}

\begin{figure}[htbp]
    \centering
    \resizebox{1.0\linewidth}{!}{\usetikzlibrary{arrows.meta}

\begin{tikzpicture}[x=1.3cm, y=1.3cm, font=\footnotesize]

  \foreach \x/\val in {1/1, 2/3, 3/4, 4/6, 5/7} {
    \draw[line width=1.3pt] (\x,-1.0) -- (\x,2.5);
    \node[draw=red, fill=red!15, circle, inner sep=2pt] (n\val) at (\x,2.7) {\val};
}


  \draw[->, line width=1.3pt, >=Stealth] (1,1.6) -- (-0.5,1.6); 

  \draw[->, line width=1.3pt, >=Stealth] (3,0.7) -- (-0.5, 0.7); 

  \draw[->, line width=1.3pt, >=Stealth] (5, -0.2) -- (-0.5, -0.2); 

  \node at (0.25,1.4) {2};
  \node at (0.25, 0.5) {5};
  \node at (0.25,-0.4) {8};

    
  \draw[->, line width=1pt, color=blue!50, >=Stealth] (-0.5, -1.3) -- (5.5, -1.3); 
  \node at (5.3,-1.55) {\textsf{Time}};

  \path[draw=none] (-1.0, -1.9) rectangle (5.8, 3);

\end{tikzpicture}}
    \caption{This is the partial order deduced from the tapped bins of (2,3,5,6,8) (alternatively, the missing codes of (1,4,7)) on a CARRY8 cell.  The time axis is as drawn in the blue arrow.  Bins expressed with black arrows do not have any restrictions on how early they can come. }
    \label{fig:DAG_eg}
\end{figure}
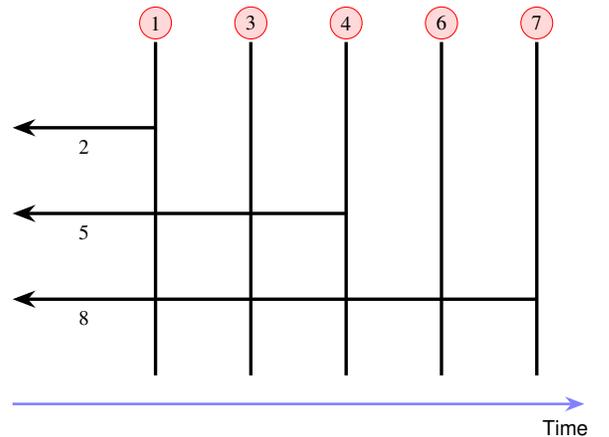

From a code density test, such partial orders can be extracted.  Logically, this is as far as one can get from the code density test.  Bins that precede the bridges are unconstrained and could occur in many permutations.  Applying this to a full TDL of hundreds or thousands of bins becomes computationally infeasible without decomposition.  The TDL clearly needs to be partitioned into certain units.  These units must be chosen such that their internal bin orderings are independent - i.e. no mixing occurs between units.  If bins from different units interfere, the number of valid permutations increases drastically, and more importantly, it becomes unclear how many units the interference spans.  A good unit division with no mixing should produce a code density test in which the final bin of each unit is tapped.  A natural unit is a CARRY block (the CARRY8 block, in this design), but empirical results show mixing across the cell boundaries. 
Therefore, we group the cells using a $\mathbb{Z}_3$ grouping scheme — that is, a single group only treats one in every three CARRY cells as a relevant unit (see Figure~\ref{fig:TDC_design}(b)) — to ensure no cross-boundary interference.  The bin sequence calibration is performed on each of the three groups individually, after which the results must be merged by ITI (see Section ~\ref{subsec:ITI}).  It is worth emphasizing that the choice of unit is, in principle, arbitrary — provided that the designated units do not interfere with one another. Larger units may be chosen to reduce the number of groups and simplify processing, albeit at the potential cost of increased permutation complexity.  Each unit’s missing code pattern defines a partial order, which we represent as a directed acyclic graph (DAG). The DAG is extracted using the algorithm in Alg.~\ref{alg1}, and an example is shown in Fig.~\ref{fig:DAG_eg}. Given the DAG, all compatible bin orderings (i.e., permutations consistent with the partial order) can be generated.

\begin{algorithm}[htbp]
\caption{Getting DAG for a CARRY8 cell}
\label{alg1}
\begin{algorithmic}
\Require tapped\_bins $\subseteq \{1,\ldots,8\}$
\Ensure DAG, in\_degree, zero\_degrees 

\State DAG $\gets$ empty map from node to another node
\State bridge $\gets \min(\text{tapped\_bins})$

\vspace{0.7ex}

\For{number = 2 to bridge}
    \State DAG[number] $\gets$ DAG[number] $\cup \{1\}$
\EndFor

\vspace{0.7ex}

\If{bridge $\ne$ 8}
    \State DAG[1] $\gets$ DAG[1] $\cup \{\text{bridge} + 1\}$
\EndIf

\vspace{0.7ex}

\State bridge $\gets$ bridge + 1

\For{value = bridge + 1 to 8}
    \If{value $-$ 1 $\in$ tapped\_bins}
        \State DAG[bridge] $\gets$ DAG[bridge] $\cup \{\text{value}\}$
        \State bridge $\gets$ value
    \Else
        \State DAG[value] $\gets$ DAG[value] $\cup \{\text{bridge}\}$
    \EndIf
\EndFor

\vspace{0.7ex}

\State in\_degree $\gets$ map from node 1 to 8 $\to$ 0
\ForAll{edge\_list in DAG}
    \ForAll{destination in edge\_list}
        \State in\_degree[destination] $\gets$ in\_degree[destination] + 1
    \EndFor
\EndFor

\vspace{0.7ex}

\State zero\_degrees $\gets$ $\left\{ \text{node} \in \{1,\ldots,8\} \mid \text{in\_degree[node]} = 0 \right\}$

\State \Return DAG, in\_degree, zero\_degrees
\end{algorithmic}
\end{algorithm}

Thanks to the FPGA's relatively uniform architecture, the missing code pattern tends to be consistent across units. In practice, only a small number of unique missing code patterns require correction, even when handling over a hundred units.  For example, empirically speaking, for around 400 time bins, there are usually only about 5 missing code CARRY8 patterns.  This also allows us to define an ansatz — a likely starting permutation — based on empirical experience.  

To guarantee accurate sequence calibration, we construct an error library. This library records, for each possible permutation of a unit, the tapped bin pattern that would result if that permutation were the true underlying sequence, with the ansatz-based guess being the perceived sequence.  Given the uniformity of tapped/missing code patterns across CARRY cells, this computation remains tractable even at the scale of a full TDL.  After performing the initial bin sequence correction and another following code density test, we use the error library to eliminate a large number of permutations that are incompatible at this stage.  In some cases, multiple permutations may yield identical tapped bin patterns; in such scenarios, we build a second-stage error library restricted to these candidates.  

We refer to this overall approach as Partial Order Reconstruction (POR).  The first POR constructs a DAG from the code density test result and yields the permutation that most closely matches the initial ansatz. From the second POR onward, calibration proceeds by traversing the error library using subsequent code density test results.  After each POR stage, FPGA resynthesis is required.  The results of POR are outlined in Section~\ref{subsec:POR_results}.

\subsection{Fine Resolution: \\ Iterative Time-bin Interleaf (ITI) \label{subsec:ITI}}
We propose another novel technique — Iterative Time-bin Interleaf (ITI) — to increase the number of time bins and enhance time resolution in FPGA-based TDCs.  Conceptually similar to multi-chain averaging, ITI also leverages multiple TDLs. However, instead of averaging fine time values across chains, ITI physically interleaves bins from multiple TDLs into a single sequence.  Specifically, it constructs a unified TDL by directly connecting every individual bin from different TDLs to a single priority encoder (see Figure \ref{fig:TDC_design}(c)). This architecture enables a true temporal ordering of bins based on physical delay differences, eliminating the need for averaging or post-processing and making the structure amenable to iterative application. 

To perform ITI, the start time of each bin can be estimated—up to the statistical resolution of the code density test—via the recursive expression:

\begin{equation}
    t[n] = \sum_{k=0}^{n-1}W[k]
\end{equation}

With $t[0] = 0$, and $W[k]$ is the bin width of $k^{\textbf{th}}$ time bin.  This definition differs slightly from the conventional bin-by-bin width calibration~\cite{sensors, liu2015tdc} in that it omits the current bin’s width in the summation.  This is required, since the starting time of the bins - independent of their widths - must be calibrated to avoid missing codes. 
 Once each bin’s start time is determined, all bins across TDLs can be globally sorted and aligned to produce a fully calibrated, sequence-consistent TDL.  





If desired, the sequence-calibrated TDL may serve as a foundation for further enhancement using complementary techniques. Alternatively, a second round of ITI can be performed to achieve even finer resolution. Although conceptually this would involve iteratively ITI-ing previously interleaved TDLs, a more efficient approach is to directly interleave all individual segments into a single TDL without intermediate grouping.  The results of ITI are outlined in Section~\ref{subsec:ITI_results}.

\section{Experimental Results, \label{sec:Exp_results}}
The experimental results of POR, ITI, bin-width calibration, and Time Interval measurements are presented in this section. The source code implementing the POR and ITI techniques, along with relevant data processing scripts, is available at the project’s GitHub repository.\footnote{See: \url{https://github.com/psunwoo/TDC_Calibration}}

\subsection{POR Result, \label{subsec:POR_results}}
\begin{figure}[htbp]
  \centering
  \includegraphics[width=0.95\linewidth]{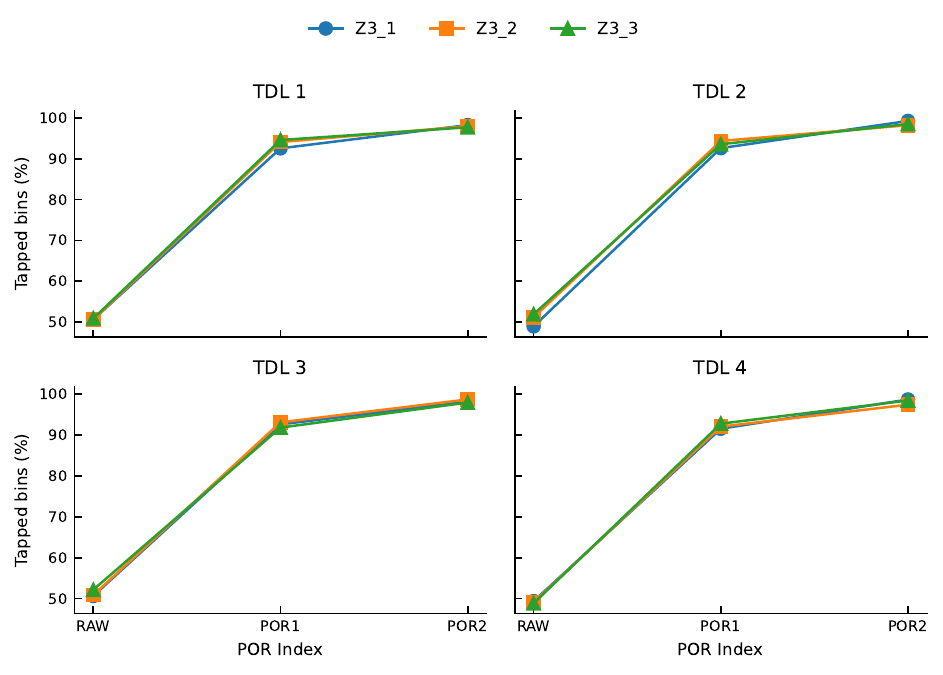}
  \caption{The percentage of tapped bins in each of the 12 segments by using POR for bin sequence calibration.}
  \label{fig:object-measurements}
\end{figure}

As can be seen in Fig.~\ref{fig:object-measurements}, in this study, the percentage of tapped bins per TDL $Z_3$ segment prior to POR calibration ranged from $48.85\%$ to $52.16\%$. Following the applications of POR twice, this metric increased significantly, reaching $97.45\%$ to $99.23\%$, across segments containing between 387 and 395 bins. After traversing on the error library once (POR2), all permutation candidates had been fully exhausted.

The fact that the post-POR2 result does not reach an exact $100\%$ is attributed to two practical factors. First, although rare, the presence of ultra-narrow bins introduces ambiguity: their widths are so small that a large volume of code density testing is required to reliably distinguish them from true missing codes.  However, when merging the corrected bin sequences in Section~\ref{subsec:ITI}, these ultra-narrow bins are intentionally filtered out to improve linearity and ensure no missing codes remain — making their exclusion acceptable in practice.  Second, minor inconsistencies in the synthesized TDL logic — particularly related to offset configurations introduced by incremental synthesis — may have introduced slight distortions. Nevertheless, under ideal conditions, the POR2 step is theoretically capable of achieving complete recovery of the tapped bin pattern.

\subsection{ITI Result, \label{subsec:ITI_results}}

\begin{figure}[htbp]
  \centering
  \includegraphics[width=0.95\linewidth]{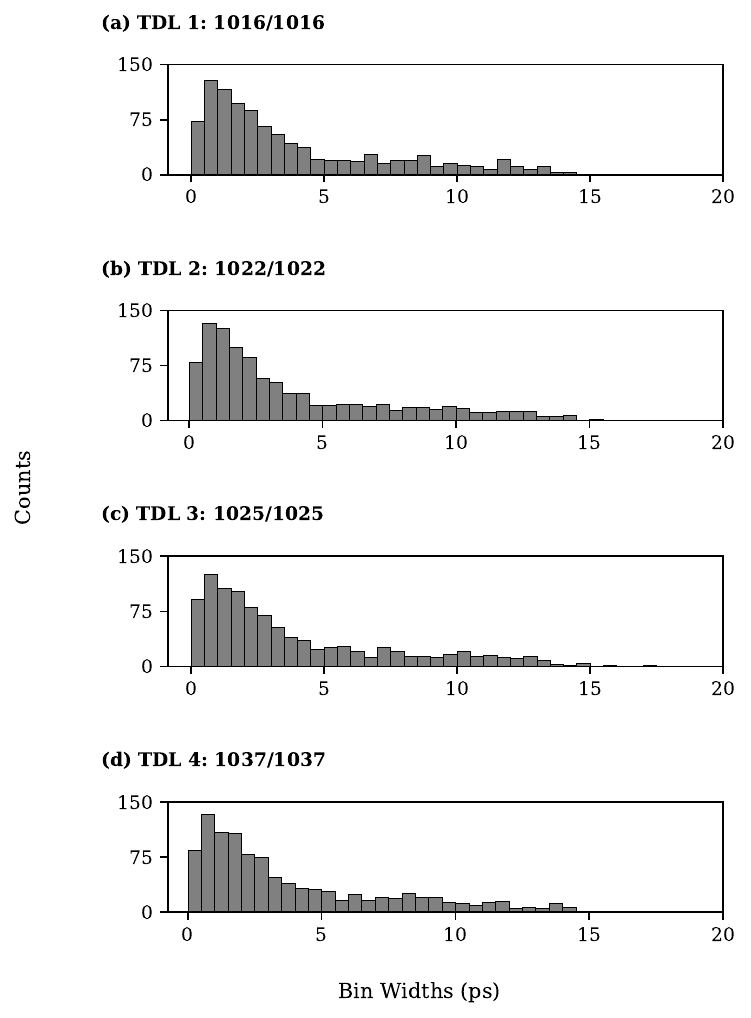}
  \caption{
Post-ITI bin width histograms for four TDLs reconstructed from bin-calibrated $\mathbf{Z}_3$ segments. Time bins with widths smaller than 0.2\,ps were filtered out during the ITI process. Each subfigure shows the number of tapped bins relative to the total number of assigned bins (e.g. 1016/1016).  Notably, ITI preserves the calibrated bin sequence and introduces no new missing codes, demonstrating its reliability as a non-destructive technique within the 0.2\,ps filtering threshold.}

  \label{fig:first_ITI}
\end{figure}

ITI is applied to the calibrated $\mathbf{Z}_3$ segments from Section~\ref{subsec:POR}, and the bin widths of resulting individual merged TDLs are presented in Fig.~\ref{fig:first_ITI}.

\begin{figure}[htbp]
  \centering
  \includegraphics[width=0.95\linewidth]{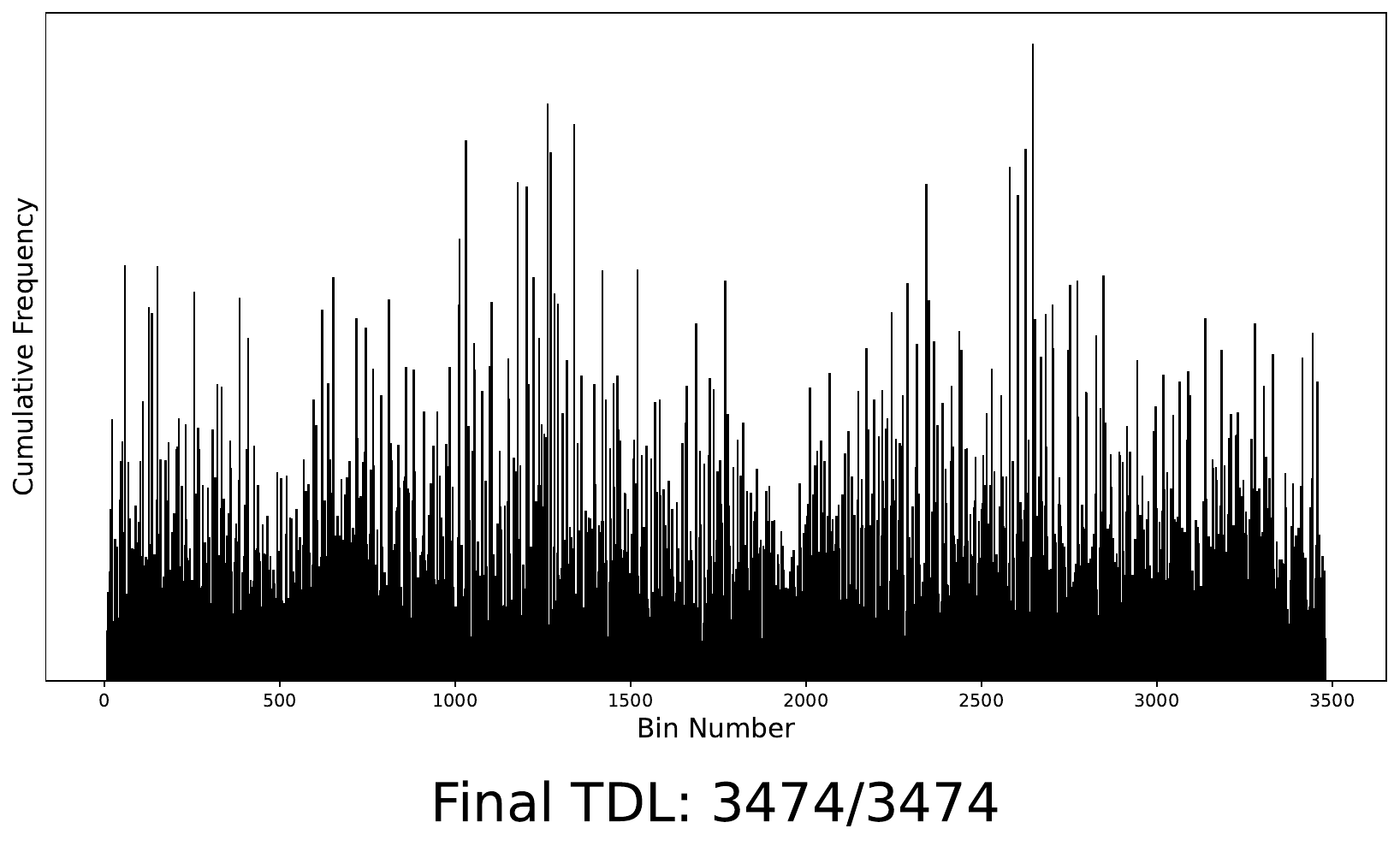}
  \caption{Final ITI result after interleaving all time bins from the bin-calibrated $\mathbf{Z}_3$ segments with 0.2\,ps filtering threshold.  All assigned bins are successfully tapped, demonstrating the effectiveness of the ITI procedure in preserving and merging the calibrated structure. }

  \label{fig:ITI}
\end{figure}

The final results merging every time bin across the 4 TDLs via ITI are presented in Fig.~\ref{fig:ITI}.  Notably, ITI improved the TDC’s resolution by a factor of 3.4 ($< 4$ due to ultra-narrow bin filtering) due to the increased bin count.  Combined with POR calibration, it results in another four-fold improvement in the number of usable time bins compared to when attempted to interleaf based solely on timing reports from Xilinx FPGA tools.  ITI is thus introduced and demonstrated in this research to be effective in 16nm FPGA with 0.2ps filtering threshold applied. This threshold was chosen as a proof-of-concept and further research could follow on the ITI threshold versus temperature and voltage variations. 

\subsection{Bin Width Calibration and Nonlinearity, \label{subsec:Bin_width_calibration}}

As shown in Figure~\ref{fig:ITI}, the bin widths deviate significantly from linearity. Nonlinearity is a critical figure of merit for TDCs, and the inherent nonlinearity arising from FPGA-based TDL implementations must be corrected through post-processing.  These nonlinearities are quantified as:

\begin{align}
\mathrm{DNL}[i] &= \frac{W[i] - \mathrm{LSB}}{\mathrm{LSB}} \\
\mathrm{INL}[i] &= \sum_{k=0}^i \mathrm{DNL}[k]
\end{align}

where $\mathrm{LSB}$ denotes the ideal bin width, given by $\displaystyle \frac{T_{\mathrm{clock}}}{\text{\# of bins}}$.

Given the absence of missing codes and the 1 ps LSB resolution in this work — compared to external jitter on the order of 10 ps — bin width calibration~\cite{embedded_bin_width_cal,Sub-TDL1,sensors} can be performed by simply scaling the count of each bin using predetermined weights. Equivalently, this can be interpreted as redistributing One-Hot Code (OHC) counts between tapped bins according to pre-calculated ratios~\cite{Sub-TDL1}.  This calibration is essential for recovering the true fine time distribution in single-shot experiments. Without it, the non-linearity distorts the distribution: larger time bins inherently accumulate more counts, even if the underlying distribution is sparse in those regions. The necessary weight factors are determined using a code density test, all of which can be finite, given no missing codes.  However as the number of time bins increases, so too does the susceptibility to statistical error. A large dataset is therefore required to accurately resolve bin widths and determine reliable weight factors.  In this work, approximately 7 GB of binary OHC data was collected to construct precise weight factors. These factors can be computed via:

\begin{equation}
    \nu_i = \frac{\mathrm{LSB}}{W[i]} = \frac{1}{\mathrm{DNL}[i] + 1}
\end{equation}

\begin{figure}[!htbp]
  \centering
  \includegraphics[width=0.95\linewidth]{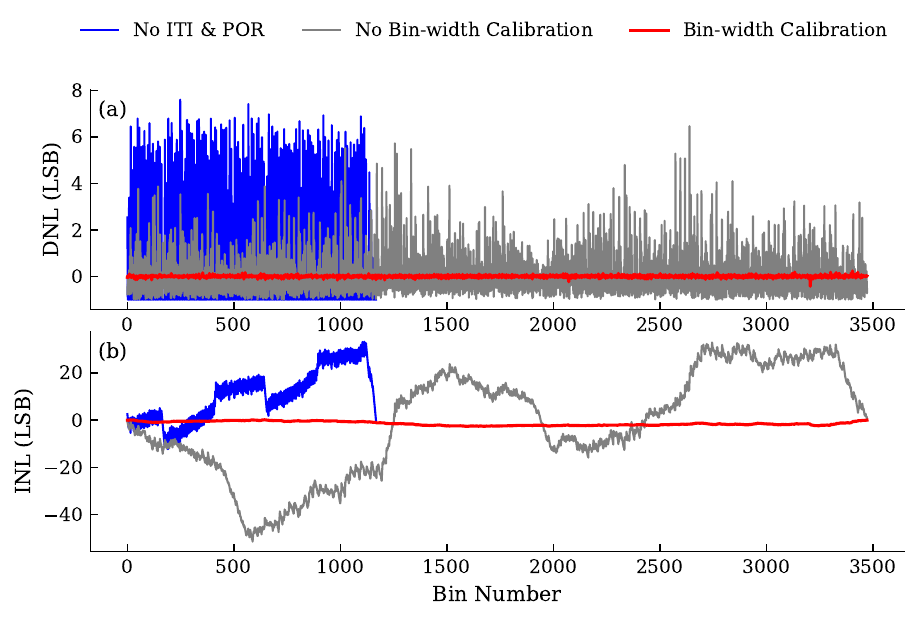}
  \vspace{2em}
  \includegraphics[width=0.95\linewidth]{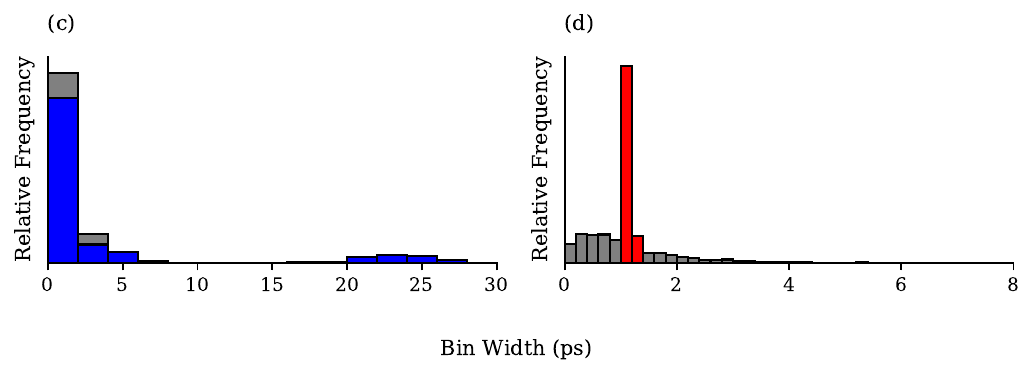}
  \caption{Blue: A single raw TDL with no ITI, POR, and bin-width calibration. Gray: A TDL with ITI and POR but with no bin-width calibration. Red: TDL in Gray after bin-width calibration. (a) DNL plots, (b) INL plots. Relative frequencies against the bin widths are plotted for Blue and Gray in (c) and Gray and Red in (d).  For quantitative results, see Table~\ref{tab:linearity}. }
  \label{fig:NL&Bin_widths}
\end{figure}

The results of this bin-width calibration are presented in Table~\ref{tab:linearity} and Figure~\ref{fig:NL&Bin_widths}.  $\omega_{\sigma}$ and $\sigma_{EQ}$ considered in Table~\ref{tab:linearity} are the equivalent bin width and the equivalent standard deviation, respectively, as studied by~\cite{Multi-chain1},~\cite{equivalent_width}:

\begin{align}
    \sigma^2_{eq} &= \sum_i \left( \frac{W[i]^2}{12} \times \frac
    {W[i]}{W_{total}}\right) \\
    \omega_{eq} &= \sigma_{eq}\sqrt{12} = \sqrt{\sum_i\left( \frac
{W[i]^3}{W_{total}}\right)}
\end{align}
where $W_{total} = \sum_iW[i]$

\begin{table}[!htbp]
\centering
\caption{Linearity performance between (1) the raw TDL, (2) TDL with POR and ITI, (3) TDL with bin width calibration from (2).}
\label{tab:linearity}
\resizebox{\linewidth}{!}{%
\renewcommand{\arraystretch}{1.3}
\begin{tabular}{l||c||c||c}
\hline
\hline

\textbf{} & 
\makecell{\textbf{No POR \& ITI} \\ \textbf{(No Bin-width Cali)}} & 
\makecell{\textbf{POR + ITI} \\ \textbf{(No Bin-width Cali)}} & 
\makecell{\textbf{POR + ITI} \\ \textbf{(Bin-width Cali)}} \\
\hline

DNL (LSB) & [--1.00, 7.60] & [--1.00, 6.46] &  [--0.43, 0.24]\\
DNL$_\text{pk-pk}$ (LSB) & 8.60 & 7.46 & 0.67\\
$\sigma_{\text{DNL}}$ (LSB) & 2.17 & 0.89 & 0.04 \\

INL (LSB) & [--12.32, 33.18] & [--51.40, 32.82] & [--2.67, 0.15]\\
INL$_\text{pk-pk}$ (LSB) & 45.50 & 84.22 & 2.82\\
$\sigma_{\text{INL}}$ (LSB) & 11.03 & 22.07 & 0.83\\

$w_{\text{eq}}$ (ps) & 6.09 & 0.73 & 0.33 \\
$\sigma_{\text{eq}}$ (ps) & 21.10 &  2.53 & 1.15 \\
Resolution (ps) & 8.40 & 1.15 & 1.15 \\

\hline
\hline

\end{tabular}%
}
\end{table}






\begin{table*}[!htbp]
  \caption{Comparison of Recent FPGA-Based TDCs}
  \label{tab:tdc_comparison}
  \centering
  \scriptsize  
  \setlength{\tabcolsep}{4pt}  

  \begin{tabularx}{\textwidth}{l l c l c c c c}
    \toprule
    \textbf{Authors} & \textbf{Methods} & \textbf{Year} & \textbf{Devices} & \textbf{LSB (ps)} & \textbf{RMS Accuracy(ps)} & \textbf{DNL (LSB)} & \textbf{INL (LSB)} \\
    \midrule

    J.Wu~\cite{Wu:2008zzr}     & WU-A & 2008 & Cyclone II & 30 & 25 & - & - \\
    J.Wu~\cite{Wu:2008zzr}     & WU-B & 2008 & Cyclone II & 2.44 & 10 & - & - \\

    H.Chen~\cite{embedded_bin_width_cal}     & Tuned-TDL, direct-histogram, multi-phase, Bin-width Cali & 2017 & Virtex 7 & 10.5 & 4.42 & [--0.04, 0.05] & [-0.09, 0.04] \\
    
    H.Chen~\cite{Sub-TDL1}     & Sub-TDL, histogram compensation, Bin-width Cali & 2019 & Virtex 7 & 10.54 & 14.59 & [--0.05, 0.08] & [-0.09, 0.11] \\ 

    H.Chen~\cite{Sub-TDL1}     & Sub-TDL, histogram compensation, Bin-width Cali & 2019 & Ultrascale & 5.02 & 7.8 & [--0.12, 0.11] & [-0.18, 0.46] \\ 

    Y.Wang~\cite{Wang:2019clu}    & WU & 2019 & Kintex 7 & 1.77 & $\sim$  2.2 & [--1, 4.3] & [-37.7, 13] \\

    P.Kwiatkowski~\cite{Kwia_2}     & Multiple TDC & 2020 & Kintex 7 & 1.01 & $<$ 3.2 & [--0.98, 2.73] & [-17.83, 5.6] \\ 

    W.Xie~\cite{Wujun_Xie}     & WU-A, Sub-TDL, Dual-sampling & 2021 & Ultrascale & 1.23 & 5.19 & [--0.84, 7.93] & [-6.36, 24.70] \\

    P.Kwiatkowski~\cite{KWIATKOWSKI2023112510}     & Multisampling, WU-B & 2023 & Kintex 7 & 0.40 & $<$3.7 & [--0.97, 5.95] & [--8.02, 219.30] \\

    Y.Wang~\cite{fine_high_DL}     & WU-A, Sub-TDL, Dual-sampling, Bidirectional Encoder & 2023 & Ultrascale+ MPSoC & 0.46 & $\sim$  2 & [--1, 6.61] & [-6.68, 62.05] \\

    Y.Zhou~\cite{zhou2023foldingtdc}     & Folding TDL & 2023 & Kintex 7 & 4.4 & $\sim$ 3.3 & [--0.9, 2.9] & [-1, 23] \\

    M.M\'{s}cichowski~\cite{MSCICHOWSKI2025115523}     & Hybrid ALM-DSP & 2025 & Intel Arria 10 & 2.1 & $<$6.36 & [--0.98, 6.01] & [14.69, 26.09] \\

    \midrule
    \textbf{This work} & POR, ITI, Bin-width Cali & 2025 & Ultrascale+ FPGA & \textbf{1.15} & \textbf{3.38} & \textbf{[--0.43, 0.24]} & \textbf{[--2.67, 0.15]} \\
    \bottomrule
  \end{tabularx}

  \vspace{1mm}
  \captionsetup{font=footnotesize}
  \caption*{\footnotesize * LSB, DNL, and INL results are averaged for the multichannel TDCs.}
\end{table*}

\subsection{Time Interval (TI) Measurements, \label{subsec:TI_results}}

\begin{figure}[!htbp]
  \centering
  \includegraphics[width=0.95\linewidth]{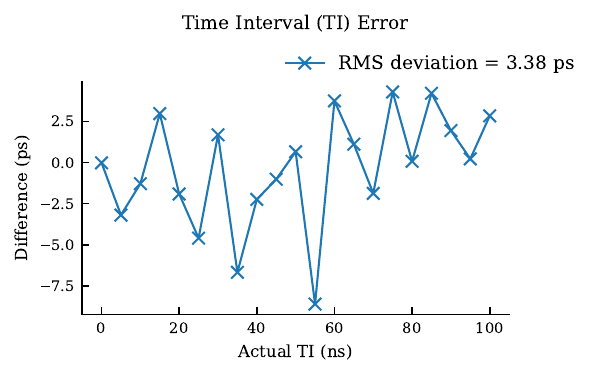}
  \caption{Measured TI deviation versus actual TI delay. Each point represents the average of three measurements. RMS precision of the TDC was $3.38$\, ps.}
  \label{fig:TI_result}
\end{figure}

On top of the bin width calibration, the standard bin time calibration to map the time bin to the fine time was carried out.  Time interval (TI) measurements were performed using pulse trains of varying time separations to evaluate the accuracy of the proposed TDC. Each data point in Fig.~\ref{fig:TI_result} represents the average of three repeated TI measurements at the specified input delay.  The performance of the purposed TDC is compared to that of other work in Table~\ref{tab:tdc_comparison}.

\section{Discussion}
In this design, the four TDLs used for ITI were implemented in parallel, which resulted in ultra-wide bins — particularly near the boundaries of the common clock domains (around bin indices 1200 and 2600 in Figure~\ref{fig:ITI}, corresponding to the three clock regions in the system) — prior to bin-width calibration.  Future implementations could mitigate this by applying phase shifts to each TDL’s clock buffer (inspired by multi-phasing techniques) or by adopting the Wave Union (WU) method to subdivide these wide bins. Subsequent application of multi-chain averaging may further enhance time resolution and precision beyond the current state-of-the-art.

In addition, employing modified encoders — such as the bidirectional priority encoder~\cite{fine_high_DL} — would allow POR to be executed in parallel across multiple segments, significantly reducing calibration time. While the proposed POR and ITI methods are currently performed offline and require FPGA resynthesis, a parallelized or partially in-situ implementation would greatly improve their practicality.

\section{Conclusion}

\begin{figure}[!htbp]
  \centering
  \includegraphics[width=0.95\linewidth]{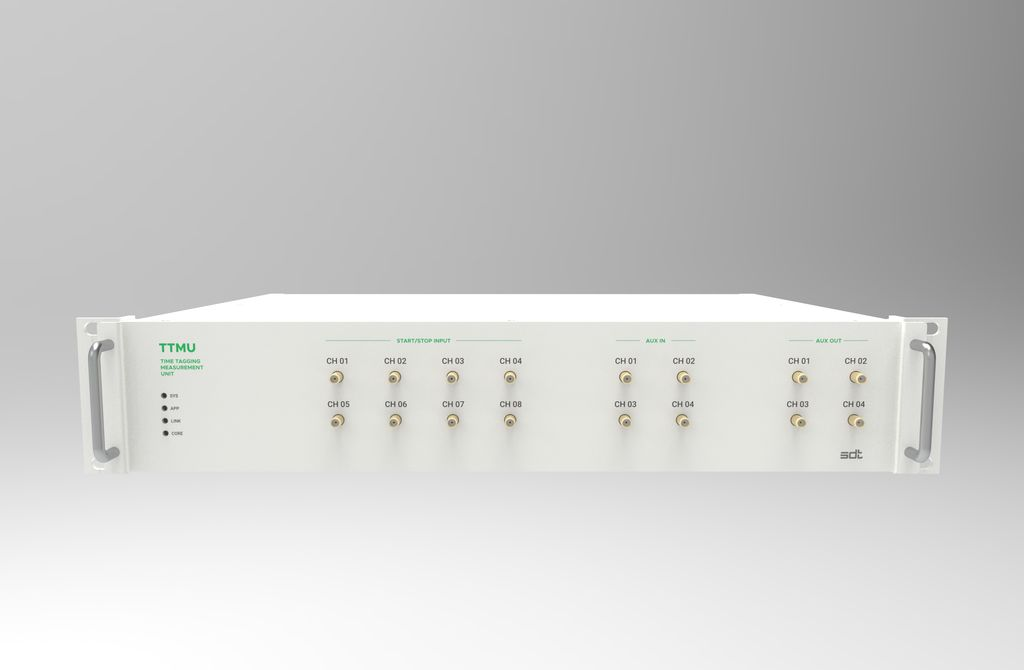}
  \caption{Image of the SDT's TDC electronics used in this work (called 'Time Tagging Measurement Unit')}
  \label{fig:TTMU_image}
\end{figure}

We proposed a Time-to-Digital Converter (TDC) architecture that achieves high performance with 1.15\,ps resolution and minimal hardware overhead.  Figure~\ref{fig:TTMU_image} is the electronics used in this paper.  This was made possible through two novel techniques that can be performed on any hardware — Partial Order Reconstruction (POR) and Iterative Time-bin Interleaving (ITI). The use of POR enables recovery of nearly all usable bins by resolving the partial ordering of bins from code density test data, while ITI improves time resolution by merging multiple calibrated TDLs into a unified delay line without relying on averaging.

As shown in Table~\ref{tab:linearity} and ~\ref{tab:tdc_comparison}, the interleaving mechanism in ITI could produce smaller nonlinearity even prior to bin-width calibration, compared to some cases. The proposed design achieves state-of-the-art performance with an RMS precision of 3.38,ps, a DNL of [–0.43, 0.24] LSB, and an INL of [–2.67, 0.15] LSB.

Moreover, due to the increasing use of FPGAs in diverse scientific and engineering contexts, the POR and ITI frameworks may find broad utility in enabling accurate delay characterization across a wide range of programmable platforms.

\section*{Acknowledgments}
The authors are deeply grateful to Youngjun Kang and Seungpil Lee for their significant contributions to this research, particularly in the development and engineering of technical features within SDT’s Time-to-Measurement Unit (TTMU). We are also thankful for Professor Yosep Kim at Korea University for his constructive feedback on the paper.


\bibliographystyle{IEEEtran}
\bibliography{cleaned_references}


 




\vfill

\end{document}